\documentclass[useAMS,usenatbib]{mn2e}

\usepackage{amssymb} 
\usepackage{verbatim} 
\usepackage{amsmath}

\usepackage{graphicx} 

\newcommand{\dzdt}{\frac{\Delta z}{\Delta t}}

\newcommand{\mnras}{\mbox{MNRAS}}
\newcommand{\aj}{\mbox{{AJ}}}
\newcommand{\apjl}{\mbox{{ApJL}}}
\newcommand{\prd}{\mbox{{PhRvD}}}
\newcommand{\nat}{\mbox{{Nat}}}
\newcommand{\apj}{\mbox{{ApJ}}}

\newcommand{\apjs}{\mbox{{ApJS}}}
\newcommand{\procspie}{\mbox{{Proc. SPIE}}}

\newcommand{\aap}{\mbox{{A\&A}}}

\title[LRGs in Simulations]{Luminous Red Galaxies in Simulations: 
Cosmic Chronometers?}

\author[S. M. Crawford et al.]
  {S. M. Crawford$^{1, 2}$\thanks{Email: crawford@saao.ac.za},
  A. L. Ratsimbazafy$^{3,4}$, C.~M.~Cress$^{3,4}$,
  E. A. Olivier$^{1,3}$, \newauthor S-L. Blyth$^{5}$, K. J. van der
  Heyden$^{5}$ \\ $^{1}$ South African Astronomicaly Observatory,
  P.O. Box 9, Observatory 7935, Cape Town, South Africa\\ $^{2}$
  Southern African Large Telescope, P.O. Box 9, Observatory 7935, Cape
  Town, South Africa\\ $^{3}$ Physics Department, University of
  Western Cape, Private bag X17, Cape Town 7535, South Africa \\
  $^{4}$ Centre for High Performance Computing, 15 Lower Hope St.,
  Rosebank, Cape Town 7700, South Africa \\ $^{5}$ Department of
  Astronomy, University of Cape Town, Private Bag X3, Rondebosch 7701,
  South Africa }

\date{\today}

\pagerange{\pageref{firstpage}--\pageref{lastpage}} \pubyear{2009}

\begin{document}

\label{firstpage}

\maketitle

\begin{abstract}
There have been a number of attempts to measure the expansion rate of
the universe at high redshift using Luminous Red Galaxies (LRGs) as
''chronometers''. The method generally assumes that stars in LRGs are
all formed at the same time. In this paper, we quantify the
uncertainties on the measurement of H(z) which arise when one
considers more realistic, extended star formation histories. In
selecting galaxies from the Millennium Simulation for this study, we
show that using rest-frame criteria significantly improves the
homogeneity of the sample and that $H(z)$ can be recovered to within
3\% at $z \sim 0.42$ even when extended star formation histories are
considered. We demonstrate explicitly that using Single Stellar
Populations to age-date galaxies from the semi-analytical
simulations provides insufficient accuracy for this experiment but
accurate ages are obtainable if the complex star formation histories
extracted from the simulation are used. We note, however, that
problems with SSP-fitting might be overestimated since the
semi-analytical models tend to over predict the late-time
star-formation in LRGs.  Finally, we optimize an observational
program to carry out this experiment.

\end{abstract}


\section{Introduction} \label{sec:intro}

An important constraint on cosmological models is the evolution of the
Hubble parameter, which is defined as:
\begin{equation}
H(z)=-\frac{1}{1+z} \frac{dz}{ dt} . 
\label{eq:hz}
\end{equation}
Most tests of cosmology use measures of the luminosity distance or the
angular diameter distance, which include an integral over $H(z)$, but
measuring $H(z)$ directly can potentially provide a more direct
constraint on cosmological parameters (Jimenez \& Loeb 2002).  $H(z)$
can be determined at high redshifts by measuring the time-interval,
$\Delta t$, corresponding to a redshift interval, $\Delta z$, to
obtain an approximation of the derivative in equation (1).  A number
of authors have attempted to use this method (Ferreras, Mechiorre,
\& Silk 2001, Jimenez \& Loeb 2002, Jimenez et al. 2003, Capozziello
et al. 2004, Ferreras, Melchoirre, \& Tocchini-Valentini 2003, Simon
et al. 2005, Dantas et al. 2007, Verkhodanov, Parijskij, \&
Starobinsky 2007, Samushia et al. 2009) to track the evolution of
$H(z)$ as a function of redshift, and place constraints on
cosmological parameters.   Most recently, Stern et al. (2010) extended
the studies of Jimenez et al. (2003) to a larger, more homogeneous sample
to make the measurement of H(z) from LRGs.

Jimenez \& Loeb (2002) suggested using spectroscopic age-dating of
Luminous Red Galaxies (LRGs) to measure $dz/dt$. LRGs are galaxies
selected from the Sloan Digital Sky Survey (SDSS, York et al. 2000) by
apparent magnitude to trace the evolution of a homogeneous,
volume-limited sample of red galaxies (Eisenstein et al. 2001).
Jimenez et al. (2003) used LRGs from the SDSS to measure $H_o=69 \pm
12$ km s$^{-1}$ Mpc$^{-1}$ (the Hubble parameter at z=0). Simon et
al. (2005) used the same method to find $H(z)$ as a function of
redshift for $0<z<2$. In both cases, the authors assume that LRGs are
drawn from the same parent population with the bulk of their stars
formed at high redshift and fit single-burst equivalent ages to the
galaxies.  Although there is evidence that LRGs make up a fairly
homogeneous population of galaxies consisting mostly of old
populations, Roseboom et al. (2006) have shown that this is not
completely true.  Furthermore, Barber et al. (2007) find that these
galaxies have a range of formation ages and stellar histories. Jimenez
and Loeb (2002) note that galaxies would not have formed at exactly
the same time and explore what effect this would have on one's ability
to discriminate between different models for w(z). They do not,
however, discuss the consequences of star formation being extended
over a period in each galaxy (they assume all stars form at exactly
the same time in a galaxy). This extended star formation history (SFH)
needs to be considered when one defines and measures an age for a
galaxy. In Jimenez et al (2003), the question is addressed in a little
more detail: they simulate LRG spectra and add a contribution from a
small fraction of stars formed at late times; they then age-date the
galaxies and conclude that the age "edge" is not significantly
different from the theoretical expectation. They do not attempt to
quantify what uncertainties arise as a result of the extended star
formation.

Given the importance of correct estimates of uncertainties on the
measured H(z), we explore, in this paper, the uncertainties that
arise from LRGs having extended and varied SFHs. We use LRGs
identified in the Millennium Simulation (MS, Springel et al. 2005)
since the only way to reliably determine uncertainties is in a
simulation where the SFH and input cosmology are known.  The MS is
currently one of the most detailed simulations of the evolution of
galaxies and large-scale structure and has been tuned to reflect
observational data.  Almeida et al. (2008) studied LRGs in simulations
and showed that the luminosity function and the clustering of LRGs are
reasonably well-modelled in their semi-analytical models of galaxy
evolution, which also reproduce many other observables.  The parameters
of the de Lucia et al. (2006) semi-analytical model have also been
successfully tuned to match the luminosity, colour, and morphology of
local elliptical galaxies, the more massive candidates of which would
be LRGs.  

On the other hand, galaxies are not modelled perfectly in the
simulations: there are indications that model galaxies assemble later
than observations may suggest (Collins et al. 2009) but more reliable
estimates of stellar mass are required.  It should be noted that these
models overestimate the small amount of evolution seen in the
properties of LRGs (Wake et al. 2006, 2008) and likely over predict
the amount of growth occurring in these sources (Masjedi et
al. 2008). Specifically for LRGs, Barber et al. (2007) find similar
star formation rates, but younger ages than the de Lucia models.  In
addition, properties of local galaxies can be reproduced with
different models for galaxy formation (e.g. Kere{\v s} et al. 2009).
However, the simulations provide viable models of extended SFHs,
allowing us to quantify associated uncertainties and optimize analysis
methods before applying them to observational data.

In this paper, we optimize the selection of LRGs based on their
properties in the MS and examine how well $H(z)$ can be recovered from
their mass-weighted age.  To explore and quantify the uncertainties
associated with extended SFHs, we then investigate how well $H(z)$ can
be determined, if the mass-weighted age of these LRGs were measurable,
using three different methods.  In practice, mass-weighted ages are
not directly measurable and accurate age-dating using LRG spectra is
potentially a bigger challenge for the cosmic chronometer approach.
We carry out an initial study of age-dating, and its implications for
$H(z)$ measurements, using the stellar population models of Bruzual \&
Charlot (BC2003, 2003) to synthesize spectra. In contrast to fitting
only single stellar population models (SSPs) (e.g. Simon et
al. 2005) or dual-burst population models (Jimenez et al. (2003)) to
determine the LRG spectral ages, we make a comparison of SSP results
to those using more complicated SFHs.  In a companion paper (Olivier
et al. 2009), we delve much deeper into issues of age-dating galaxy
spectra.

The outline of the paper is as follows: In \S \ref{sec:sims} we
discuss the extraction of LRGs from the MS. In \S\ref{sec:cosmo} we
investigate how well $H(z)$ is recovered in the MS using LRGs
identified in the simulation. In \S \ref{sec:model}, we describe the
modelling and age-dating of LRG spectra. In \S\ref{sec:obs}, we
discuss an observational program to measure $H(z)$. We summarize our
results and conclusions in \S \ref{sec:conc}.  Throughout this work,
we adopt the cosmology used in de Lucia et al. (2006) of
$\Omega_m=0.25$, $\Omega_{\Lambda}=0.75$, and $h=0.73$.

\section{Identifying LRGs in the Millennium Simulation}\label{sec:sims}

The Millennium Simulation is a dark-matter-only N-body simulation run
using the Gadget code of Springel et al. (2005). The simulation has
$~10^{10}$ particles distributed in a 500$h^{-1}$ Mpc$^3$ box. Dark
matter halos are identified in snapshots of the simulation at 63
different times and merger trees of halos are constructed.
Semi-analytical modelling of galaxy evolution is carried out using the
merger trees, providing a catalogue of galaxies in each snapshot,
where each galaxy has an array of observational properties associated
with it (Baugh et al. 2005, Bower et al. 2006, de Lucia et
al. 2006). Properties of the halos and simulated galaxies are stored
in a database which can be accessed from the web.

Here, we focus on the galaxy catalogues constructed using the
semi-analytical models of de Lucia et al. (2006). We identify
simulated galaxies with LRG properties in various snapshots
(corresponding to a particular set of redshifts near $z\sim 0.5$) and
then extract the star formation histories. The model does
reproduce many properties of local galaxies and LRGs (Croton et
al. 2006, de Lucia et al. 2006), but a number of weakness are apparent
as outlined in the introduction.

In addition to the de Lucia models, we examined the models by
Bower et al. (2006, hereafter Durham models), which also uses the
Millennium Simulation.  However, the two models are different in the
implementation of their semi-analytical models.  For our purposes, the
most relevant difference between the models is in their handling of
gas cooling, star formation, and feedback as these are directly
related to the star formation histories of the galaxies.   We save
the comparison of other semi-analytical models and how they would
affect the calculation of $H(z)$ for future work.

\subsection{Selecting LRGs using SDSS criteria}\label{sec:sdsssel}

\begin{figure}
  \includegraphics[width=8.5cm,keepaspectratio]{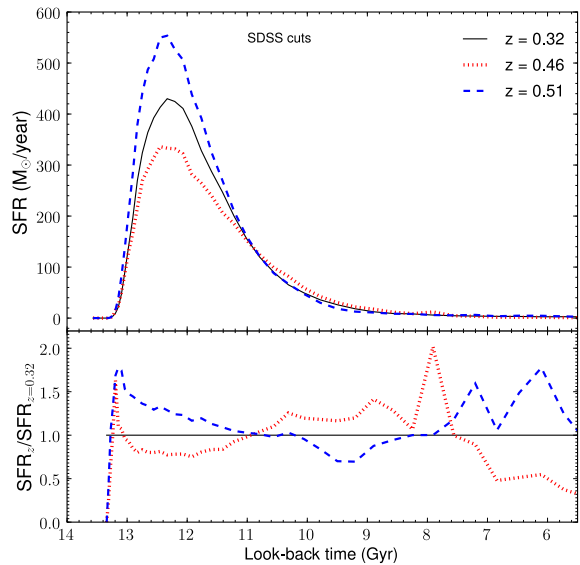}

  \caption{\textbf{Top:} The average star formation histories
  (star formation rate vs. look-back time from today) for LRGs selected
  from the MS using the SDSS cuts in Eisenstein et al. (2001). The
  solid line is for LRGs at $z=0.32$, the dotted line is for $z=0.46$
  and the dashed line is for $z=0.51$. \textbf{Bottom:} The ratio of
  the average star formation histories in the top panel to the average
  star formation history for the $z=0.32$ sample. The line-styles
  represent the same redshift bins as for the top panel. }
  
  \label{fig:sdss}
\end{figure}

Initially we extracted LRGs using the apparent magnitude constraints
for SDSS data from Eisenstein et al. (2001). Our target redshift is
close to the redshift where the SDSS `Cut I' must be replaced with the
`Cut II' colour selection criteria (At $z \sim 0.4$, the Balmer break
moves into the r-band and a different colour selection is required to
identify LRGs).  In Fig. \ref{fig:sdss}, we plot the average SFH for
three redshift snapshots at $z=0.32, 0.46, 0.51$.  The SFHs,
especially once normalized to the SFH at $z=0.32$, show significant
differences.  Importantly, there is no monotonic behaviour for
successive $z$.  In Table 1, we summarize the average formation
parameters for the population.  There is a significant dispersion in
formation redshift ($z_f$), the redshift of peak star formation
($z_p$), and the redshift of last major star formation epoch ($z_e$),
which is defined as the redshift where the star formation rate falls
below 10 $M_{\odot} \ yr^{-1}$.  For galaxies in the MS at $z=0.51$,
almost $30\%$ have had a burst of star formation greater than 5
$M_{\odot} yr^{-1}$ in the previous gigayear and up to $19\%$ are
greater than 3$\sigma$ away from the average star formation history.

In addition, the change in the average mass-weighted age, $Age_{mw}$,
does not correspond to the change in the age of the Universe at these
redshifts.  Between $z=0.32-0.46$, the age of the simulated Universe
changes by 1.13 Gyr, whereas the change in $Age_{mw}$ corresponds to
1.27 Gyr.  For $z=0.46-0.51$, the values are 0.26 and 0.18 Gyr,
respectively. A key assumption of this technique is that the galaxies
are passively evolving.  If true, then the change in their age with
redshift should corresponding to the change in the age of the
Universe.  However, that is not the case for this sample.

The sample selected using the SDSS cuts in the MS are not very
homogeneous. A source of this inhomogeneity is the selection of
galaxies in apparent magnitude space.  Due to the width of the
redshift bin within which the apparent colour cuts are defined,
different types of galaxies are selected near the edges of the
redshift bin including a large number of galaxies that have more
extended SFHs.  Inclusion of these galaxies decreases the sensitivity
of this method for determining $H(z)$.  Furthermore, the observing
efficiency is reduced if only the oldest galaxies are used unless they
can be pre-selected.

\begin{table*}
  \centering
  \begin{minipage}{140mm}
  \caption{Formation Properties of LRGs}
  \begin{tabular}{rlllllll}
\hline 
\multicolumn{1}{c}{Cut} & 
\multicolumn{1}{c}{z}  & 
\multicolumn{1}{c}{$z_f$} & 
\multicolumn{1}{c}{$z_p$} & 
\multicolumn{1}{c}{$z_e$} & 
\multicolumn{1}{c}{$Age_{mw}$} &
\multicolumn{1}{c}{$f_{sfr}$\footnote{The fraction of galaxies with $SFR > 5 \ M_{\odot} yr^{-1}$ within 1 Gyr of the redshift.}} &
\multicolumn{1}{c}{$f_{3\sigma}$} \\ \hline
SDSS & 0.32 & $11.86 \pm 1.30$ & $4.95 \pm 1.26$ & $1.46 \pm 0.49$ & $8.32\pm 0.28$ & 0.32 & 0.13  \\
SDSS & 0.46 & $11.94 \pm 1.23$ & $5.06 \pm 1.46$ & $1.44 \pm 0.44$ & $7.05\pm 0.35$ & 0.03 & 0.19  \\
SDSS & 0.51 & $12.17 \pm 1.16$ & $5.05 \pm 1.18$ & $1.49 \pm 0.47$ & $6.87\pm 0.22$ & 0.29 & 0.17  \\
Our Cut & 0.32 & $11.94 \pm 1.23$ & $4.91 \pm 1.14$ & $1.48 \pm 0.47$ & $8.29\pm0.27$ & 0.14 & 0.15 \\ 
Our Cut & 0.46 & $11.92 \pm 1.23$ & $4.96 \pm 1.18$ & $1.54 \pm 0.48$ & $7.20\pm0.29$ & 0.14 & 0.15 \\ 
Our Cut & 0.51 & $11.89 \pm 1.25$ & $5.00 \pm 1.21$ & $1.58 \pm 0.47$ & $6.83\pm0.31$ & 0.14 & 0.15 \\
\hline
   \end{tabular}
   \end{minipage}
\end{table*}

\subsection{Selecting a more homogeneous sample}\label{sec:civ}

Since the age-dating experiment assumes a similar SFH for LRGs at
different redshifts, we explored other selection criteria that would
yield a more homogeneous set of objects.  The MS includes rest-frame
luminosities, and instead of the SDSS selection, we chose to use
absolute magnitude cuts in the rest-frame to select LRG-like objects.
In a real age-dating experiment, this would require having a large
sample of galaxies with multi-band photometry and spectroscopic (or
high quality photometric) redshifts.  For $z\sim0.5$, this selection
can easily be done from the SDSS data (e.g. the 2SLAQ catalog of
Cannon et al. 2006).

\begin{figure}
  \includegraphics[width=8.5cm]{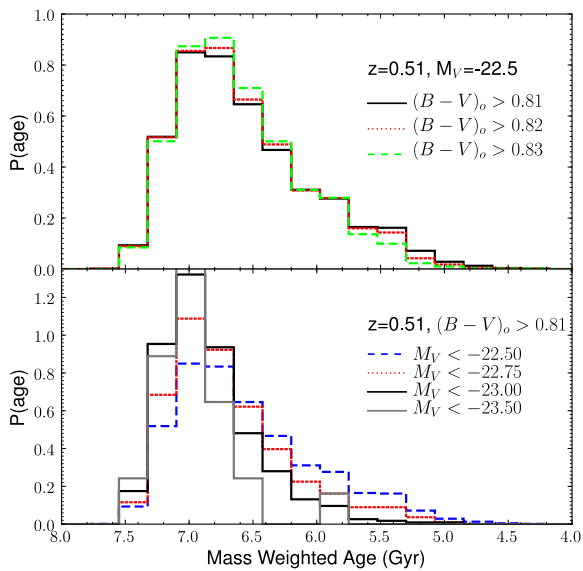} 

  \caption{The effect of varying the brightness cut \textbf{(top)} and the
  color cut \textbf{(bottom)} on the mass-weighted age histogram for galaxies
  selected at z=0.51.   }

  \label{fig:cuts}
\end{figure}

The effect of varying the colour and brightness on the normalized
distribution of mass-weighted ages of simulated galaxies is shown in
Fig. \ref{fig:cuts} for galaxies at $z=0.51$.  We demonstrate that by
selecting the brightest galaxies, we are able to select galaxies with
narrower SFHs; that is, galaxies in which more of the stars are formed
at earlier times. In the lower panel, we show the effect of changing
the colour-cut, demonstrating that the SFHs are not very sensitive to
the rest-frame colour-cut we choose. Using these results, we decided
to use a rest-frame colour-cut of $B-V>0.81$ and an absolute magnitude
cut of $M_V<-23$.  A brighter magnitude cut does produce an even more
homogeneous sample, but results in a dramatic drop in the number of
sources (a factor of 10 decrease with a 0.5 mag increase).

\begin{figure}
   \includegraphics[width=8.5cm,keepaspectratio]{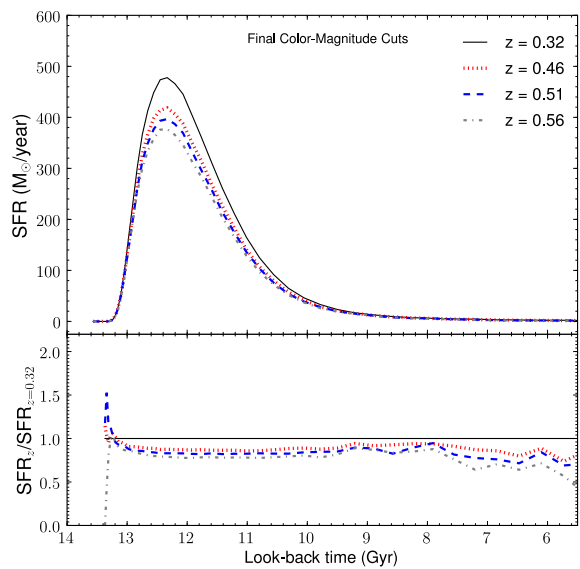}
   \caption{\textbf{Top:} The average star formation histories
   (star formation rate vs. look-back time from today) for LRGs
   selected from the MS using our revised absolute magnitude cuts. The
   solid line is for LRGs at $z=0.32$, the dotted line is for
   $z=0.46$, the dashed line is for $z=0.51$ and the dot-dashed line
   is for $z=0.56$. \textbf{Bottom:} The ratio of the average
   star formation histories in the top panel to the average
   star formation history for the $z=0.32$ sample. The line-styles
   represent the same redshift bins as for the top panel. }
 
   \label{fig:civ}
\end{figure}

\begin{figure}
  \includegraphics[width=8.5cm,keepaspectratio]{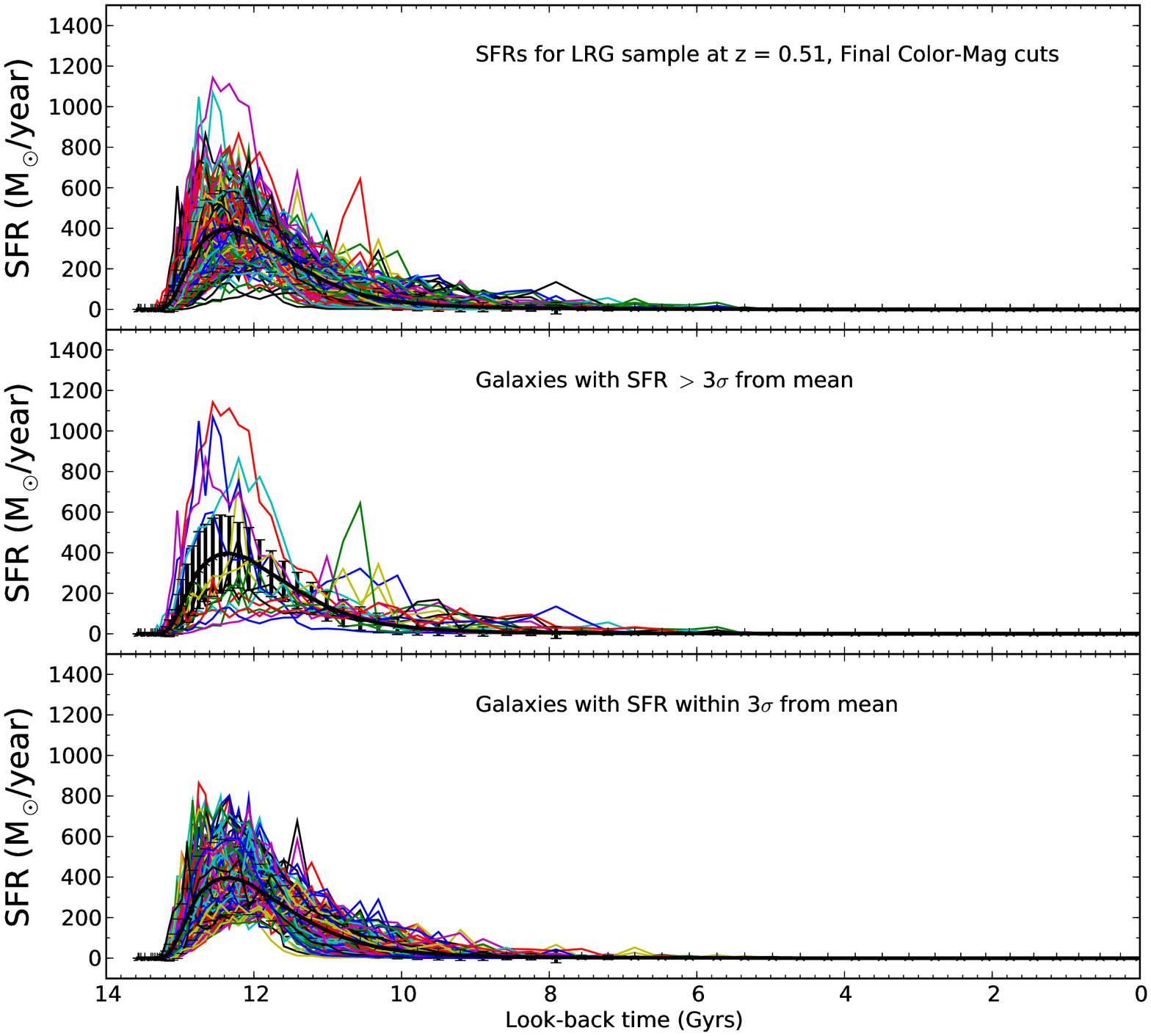}
  \caption{\textbf{Top:} Star formation histories of 200 galaxies
  selected using the revised cuts from the MS. The black line
  indicates the mean star formation history of 1440 galaxies selected
  using the cuts. The error bars on the black line represent the
  standard deviation of the mean. \textbf{Middle:} Galaxies from the
  sample in the top panel whose star formation rates differ by more
  than $3\sigma$ from the mean star formation history.
  \textbf{Bottom:} Galaxies from the sample in the top panel whose
  star formation histories differ by less than $3\sigma$ from the mean
  star formation history.}  
  
  \label{fig:varsfh}
\end{figure}

In Fig. \ref{fig:civ}, we plot the average SFHs of galaxies selected
using our rest-frame cuts at four redshifts (upper panel) as well as
the ratio of SFR to that obtained at z=0.32 (lower panel). It is
clear, as compared to Fig. \ref{fig:sdss}, that the galaxies selected
in this way have more homogeneous SFHs, and the small change from low to
high redshift is monotonic. To demonstrate the variation in SFH for
galaxies selected in this way at a particular redshift, we plot, in
Fig. \ref{fig:varsfh}, SFHs of 200 randomly chosen galaxies from the
sample at $z=0.51$. In the lower panels, we break the 100 galaxies into
two samples: those that are more than 3$\sigma$ away from the mean and
those that are within 3$\sigma$ of the mean.

The formation properties of these galaxies are compared to the SDSS
sample in Table 1.  On average, this population shows similar
dispersion to the objects selected from the SDSS cuts.  However, a
smooth and gradual change is evident for our cut as opposed to the
dramatic change seen using the SDSS cuts.  In addition, a constant and
lower percentage of objects have had significant events of star
formation in the past as opposed to the SDSS cuts.

\begin{figure}
  \begin{center}
  \includegraphics[width=8.5cm,keepaspectratio]{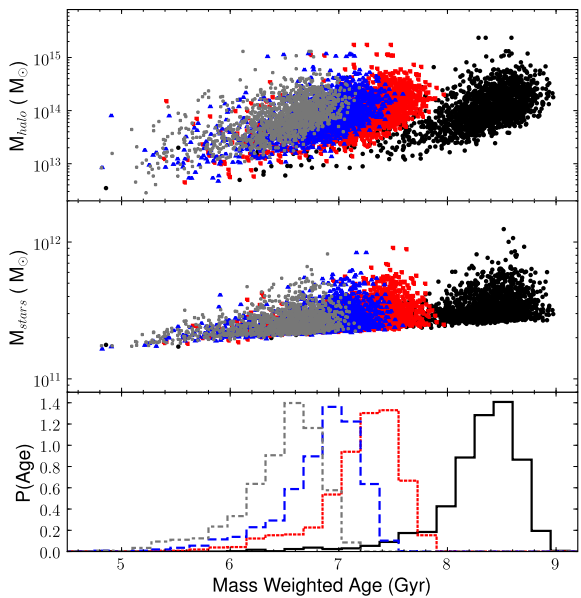}
  \caption{ \textbf{Top:} The masses of the halos hosting LRGs, plotted
  against their mass weighted ages. For all four redshifts, a strong
  correlation is found between age and halo mass.  \textbf{Bottom:} Histogram
  of the mass-weighted ages for four redshifts.}
  
  \label{fig:fig4}
  \end{center}
\end{figure}

To investigate further the properties of the LRGs selected with the
new cuts, we plot, in the top panel of Fig. \ref{fig:fig4}, the mass
of the halo which hosts the galaxies vs. the mass-weighted age of our
selected LRGs for four snapshots.  We note the more massive halos
contain older galaxies, a correlation which arises naturally in
hierarchical structure formation scenarios (de Lucia et al. 2006) and
corresponds to the observed phenomenon of downsizing (Cowie et
al. 1996).

In addition, these galaxies have large stellar masses with values
typically above $2 \times 10^{11}$ $M_{\odot}$ with a peak value just
above $10^{12}$ $M_{\odot}$.  There is a strong, lower mass cut-off in
each of the redshift bins due to the absolute magnitude cut in the
selection function.  The overall stellar mass values are in good
agreement for stellar masses calculated for LRGs by Barber et
al. (2007), but their data does not show the strong correlation
between age and stellar mass for SDSS selected LRGs that is seen in
the de Lucia models.

When we substitute metallicity for halo mass, we see no
correlation. The lower panel of Fig. \ref{fig:fig4} shows a normalized
distribution of the mass-weighted ages of galaxies from our selection.
Even at very close redshifts, the ages of the galaxies are separated
into distinct groups, although a small tail of younger galaxies is
still present.

For the rest of the paper, we refer to these galaxies as LRGs.
Galaxies selected using the old cut will be referred to as SDSS
LRGs. At four different redshift ($z=0.32, 0.46, 0.51, 0.56$), the
total number of LRGs extracted from the MS are 1705, 1491, 1448, and
1337 galaxies respectively.

\subsection{Selecting LRGs from Durham Models}

We applied both the SDSS and the absolute magnitude cuts to the Durham
model.  Although similar density of objects were found in the Durham
model at z=0.51 by using the SDSS cuts, the density of LRGs after the
absolute magnitude cut was down by an order of magnitude as compared
to both the real object density and the de Lucia model.  This was due
to the lack of bright objects in the Durham model at intermediate
redshifts.  This was previously shown by Almeida et al. (2008) in the
mismatch between the observed luminosity function of 2SLAQ LRGs and
the Durham models, where the Durham model has an excess of faint LRGs.  

For the selection of objects that were found in the Durham model, they
had an elevated level of star formation, bluer colors, less regular
stellar histories, and a large spread in the stellar ages.  These
results are consistent with results found by Almeida et al. (2008) and
Gonzalez et al. (2009) in studies of the Durham model and LRGs.
Further study is necessary to ascertain the reasons for the
differences between the models and the observations, and for this
reason, we focus on the results from the de Lucia models for the
remainder of this paper.

\section{Recovering $H(z)$ from Millennium LRGs }\label{sec:cosmo}

In this section, we explore the uncertainties on $H(z)$ related to the
extended star formation histories of LRGs and ignore difficulties
associated with measuring the age of LRGs. We use only the galaxies
selected with our rest-frame magnitude and color cuts.  We use the
mass-weighted ages, which are shown in the bottom panel of
Fig. \ref{fig:fig4}, for galaxies in the four redshift snapshots
initially used. In addition to these redshifts, we extract another set
of galaxies from snapshots with $0<z<1$, where we have adjusted our
color cuts relative to $z=0.51$ to account for passive evolution.

\begin{figure*}
  \includegraphics[width=13.0cm,keepaspectratio]{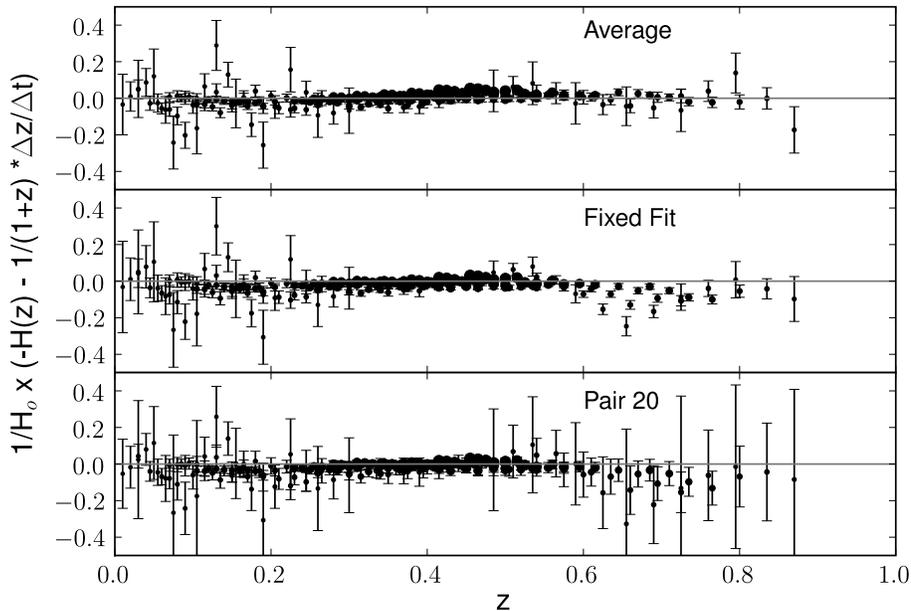} 
  \caption{The
  difference between $H(z)$ and $\dzdt$ calculated from the Average
  (top), Fixed Fitted (middle), and Pair 20 (bottom) methods. See
  the text for the details of each method.  The size of each point is
  related to the distance between the redshift snapshots, which range from
  $\Delta z=0.02-0.91$.  Larger errors are typically associated with
  smaller redshift gaps.}  \label{fig:hz}
\end{figure*}

The distribution of the ages, despite refining the cuts to provide a
narrow distribution, are still highly non-Gaussian with a tail
reaching to younger ages. With this in mind, we compare three
different methods to determine the characteristic age at a given
redshift: (1) Calculating the average of the ages, (2) fitting a
function to the distribution of ages, and (3) matching pairs of
galaxies.  For each approach, we calculate $H(z)$ using equation
(\ref{eq:hz}).  If we assume no error in the redshift, the error in
$H(z)$ will only depend on the age at the two redshift bins:
\begin{equation}\label{eq:hzerr}
\frac{\sigma_H^2}{H(z)^2} = \frac{(\sigma_{t_1}^2+\sigma_{t_2}^2)}{(t_1-t_2)^2}.
\end{equation}
In Fig. \ref{fig:hz}, we present estimates of $H(z)$ according to
various methods.  We plot the difference between our calculated value,
$\dzdt$, and the expected value for $H(z)$ based on equation (3) in
Jimenez \& Loeb (2002):
\begin{align}
-(1+z) \frac{H(z)}{H_o} = -(1+z)^{5/2} [\Omega_m(0)+ \Omega_{de}(0) \notag \\ \times \exp(3 \int_0^z \frac{dz'}{1+z'} w_{de})]^{1/2}, 
\end{align}
where $w_{de}= P_{de}/\rho_{de}$ is the equation of state parameter
for the dark energy, and in a $\Lambda CDM$ model of de Lucia et
al. (2006), $w_{de}=-1$.

The average method (top panel) recovers $H(z)$ over a range of
redshifts reasonable well.  If we assume an error in the mean age at a
specific redshift of 0.03 Gyr (this error will be discussed in \S
\ref{sec:obs}), then $H(z)$ can be calculated to a precision of
$1.6\%$ at $z\approx0.42$ (using the redshift interval between
$z_1=0.32$ and $z_2=0.51$). As can be seen in equation
(\ref{eq:hzerr}), the error in $H(z)$ goes as the inverse of the
difference in age: the closer the redshift bins, the larger the error.
Between $z=0.51$ and $z=0.56$, a small, systematic error in the mean
age of as little as $0.5\%$ will result in the calculation of $H(z)$
to be off by $10\%$.  Because of the small curvature of $H(z)$ over
these redshift ranges, large values in $\Delta z$ do not introduce a
systematic difference in the value of $H(z)$.

The middle panel presents the results for fitting the distribution of
ages to a function and determining a characteristic age.  This is akin
to fitting the envelope of oldest galaxies as in Jimenez et
al. (2003).  The data were fitted to the following function:
\begin{equation}\label{eq:pdf}
P(t) = \frac{a *g(t)^2}{1+g(t)^b}
\end{equation}
where,
\begin{equation}
(t)=\frac{t_o-t}{c}
\end{equation}
For the first test, we allowed $a$, $b$, and $c$ to be free parameters
along with $t_o$.  In the second test, we set $b=5.52$ and $c=0.66$,
which were the average parameters found during the free fits, and
solved for $a$ from the best fit value of $t_o$. The results for
$H(z)$ between the free and fixed fits were very similar and
Fig. \ref{fig:hz} (middle panel) shows the results for the fixed fit.
Unlike the average estimate, $H(z)$ is slightly underestimated at the
highest redshifts. This is probably a result of the fitting function
in equation (4) not fitting the normalized age distribution at higher
redshifts well. At $z\approx0.42$, $H(z)$ can be calculated to
$1.1\%$ by using our full sample at $z=0.32, 0.51$.

Finally, we calculate the value of $H(z)$ at a given redshift based on
the distribution of $H(z)$ calculated between each pair of galaxies in
the different redshift bins.  If we used the entire sample, overlap
between the age distributions significantly skews the results. Less
than 10 galaxies results in anomalous values of $H(z)$.  Using the 20
oldest galaxies minimizes both the systematic and random errors on the
calculation of $H(z)$, however the results are not substantially
different for using the 1000 oldest galaxies.  Once we have calculated
the value of $H(z)$ from all of the pairs, $H(z)$ at a given redshift
is the three-sigma clipped mean of the distribution.  The error is
then given by the standard deviation of the sigma-clipped
distribution. The $H(z)$ estimate at high redshifts is also slightly
below the model, but well within the errors.  At $z\approx0.42$,
$H(z)$ can be calculated to $2.8\%$

It is reassuring that we can recover $H(z)$ at a a single redshift
using various metrics for the characteristic age of the population.
Even the very gross estimate of the mean of the distribution provides
very high accuracy ($<2\%$) on the calculation of $H(z)$. Obviously
using the ages from the MS is a best case scenario where the galaxy
ages are determined without any error.  In the next section, we
explore the errors associated with age-dating LRGs.

\section{Modelling and Age-Dating LRG spectra}\label{sec:model}

There are a number of stellar population synthesis codes available
which can be used to generate synthetic spectra once a star formation
history has been determined (e.g. Fioc and Rocca-Volmerange
1997, Bruzual \& Charlot 2003).  However, the underlying physics is
not always well understood: thermally-pulsating AGB stars may not be
correctly modelled which effects infrared magnitudes significantly
(Maraston 2005); an evolving Initial Mass Function (IMF) might be
required to model cluster red-sequence galaxies which include many
LRGs (van Dokkum 2008); related to the IMF, the alpha-element
enhancement of old galaxies is not correctly modelled (Walcher et
al. 2009); and, finally, dust modelling remains uncertain.  Newer
generations of models are currently being developed and tested
(e.g. Maraston 2005, Conroy et al. 2009) along with improvements in
the underlying stellar evolution models (Marigo et al. 2008).  In
addition, there is also uncertainty in the best method for deriving
accurate ages both at low redshift (Kannappan \& Gawiser 2007, Wolf et
al. 2007, Trager \& Somerville 2009) and at high redshift (Longhetti
\& Saracco 2009, Maraston et al. 2009, Muzzin et
al. 2009). Unfortunately, the focus in these studies is more on
broad-band and low-resolution spectra rather than the high-resolution,
high-signal to noise (SNR) spectra we expect to need.

In this paper, we carry out a simple study of the age-dating question
and leave a detailed discussion for the companion paper, including a
comparison of different population synthesis models, age-dating
techniques, and comments on IMF evolution, alpha-element enhancement
and the role of AGB stars.

\subsection{Modelling LRG Spectra}\label{sec:bc}

We build synthetic spectra of LRGs by combining single stellar
population (SSP) spectral libraries of BC2003. The libraries are
tabulated at ages ranging from $10^5$ years to $2\times 10^{10}$ years
at a resolution of $\sim$3 \AA\ across the whole wavelength range from
3200 to 9500 \AA\ for a wide range of metallicities ($Z=0.0001$ to
$Z=0.05$). Spectral coverage over a larger wavelength range, from 91
\AA\ to 160 $\mu m$ is also available but at lower resolution.  For
our studies here we have assumed the standard Salpeter initial mass
function, with mass cut-offs at 0.1 and 100 $M_{\odot}$. We refer the
reader to BC2003 for details about the stellar evolution prescription
used in constructing their libraries.

From the semi-analytic models of galaxy evolution based on the MS, one
can extract both the star formation rate and metallicity of the cold
gas out of which the stars are formed, as a function of time.  Using
these and the BC2003 SSP spectral libraries as input we calculate the
emergent spectrum according to
\begin{equation}
F({\lambda}) = \int_{0}^{t_{form}}  S(t) F_{SSP}({\lambda},t,Z(t)) dt .
\label{eq:specgen}
\end{equation}
In this equation, $F({\lambda})$ is the emergent spectrum in the rest
frame; $t$, the look-back time from the redshift of interest;
$t_{form}$, the look-back time at the start of formation' $S(t)$, the
star formation rate per unit mass per unit time;
$F_{SSP}({\lambda},t,Z(t))$, the spectrum of an SSP as a function of
age; and metallicity normalized to unit mass of stars and $Z(t)$ is
the metallicity as a function of look back time.  We ignore the
effects of dust, since these LRGs are assumed to be devoid of any
remaining gas and dust (Barber et al. 2007).

We generate spectra for all 1705, 1491, 1448, and 1337 galaxies
selected from the MS at $z=0.32, 0.46, 0.51, 0.56$.  In particular, we
use the $z=0.51$ galaxies to examine the limitations in age dating the
stellar populations.

\begin{figure}
  \includegraphics[width=8.5cm,keepaspectratio]{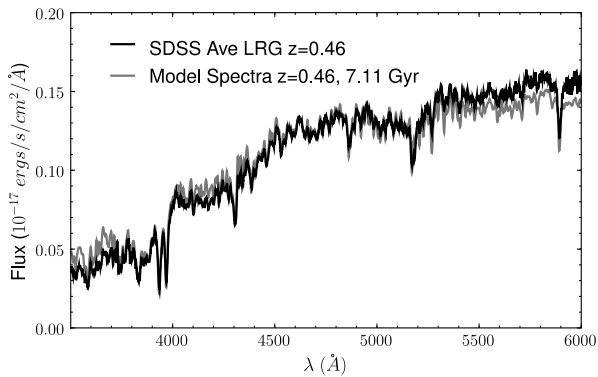}

  \caption{Average LRG spectrum at z=0.46 created by using 200
  galaxies that satisfy the revised cuts in \S \ref{sec:civ} from the
  SDSS compared to the best-fit model spectrum produced using BC2003
  with the SFH from the MS. }

  \label{fig:spectra}
\end{figure}

We used the revised colour-magnitude cuts discussed in \S
\ref{sec:civ} to extract a sample of galaxies from the SDSS
catalogue (Abazajian et al. 2009). In Fig. \ref{fig:spectra}, we
present the average spectrum of the first 200 galaxies that match our
cuts.  For comparison, we plot the best fit model spectrum at
$z=0.46$.  The fit was made from 3500-6000 \AA, which includes only
the high-resolution portion of the models.  The best fit model has a
mass-weighted age of $7.11$ Gyrs.  The overall shape of the model is
in good agreement with the observed spectrum although there are some
inconsistencies at the edges.

\subsection{Age-dating simulated LRG spectra with SSPs}\label{sec:dating}

In our first attempt to age-date simulated LRGs, we determine the
ages for the 1448 model galaxies at $z=0.51$ using the SSP template
library.  For each source, we fit the full spectrum from
$\lambda=3500-9000$ \AA\  using the SSP library described in \S
\ref{sec:model}.  We do not add any flux errors or noise to the
spectra and determine the best fit by minimizing the $\chi^2$.

For our entire sample, the systematic and random errors from fitting
the SSPs are listed in Table 2 as a function of redshift.  It is not
surprising to find a systematic error in the age determination using
SSPs as compared to the mass-weighted age as the most appropriate
comparison would be to the light-weighted age.  Trager \& Somerville
(2009) perform a similar comparison for red galaxies in the Coma
cluster comparing observed and simulated populations with SSPs.  They
also find that the SSPs underestimate the age of the galaxies and the
SSP age is poorly correlated with the mass-weighted age due to recent
star formation dominating the spectra.  We also find that the offset
in the systematic error does not correlate with the age of the
Universe.  The differential ages are less sensitive to systematic
errors (Jimenez et al. 2004), and if the systematic errors were
constant, this would not affect the calculation of $H(z)$.  Since the
bias does change with redshift, this prevents SSP-determined ages from
being very useful for determining $H(z)$.

Based on the sample from McCarthy et al. (2004) of early type
galaxies, Simon et al. (2005) claim they are well fit by a single
burst population and use these ages to recover a value of $H(z)$.
This is consistent with our results presented here as the
resolution of their spectroscopic data ($17$ \AA) and the precision on
$H(z)$ are much greater than those that we examine here.  McCarthy et
al. (2004) do fit a majority of their objects with burst of less than
0.1 Gyrs, but the objects have a range of ages and formation redshifts
that vary as much as 3 Gyrs.  This is consistent with SSP fitting
being dominated by the most recent star formation.

\begin{figure}
  \includegraphics[width=8.5cm,keepaspectratio]{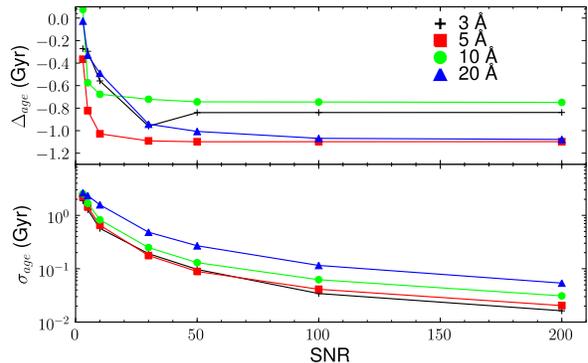}

  \caption{\textbf{Top:}  The  average systematic  offset between  the
  model age  and   the mass-weighted age  as a function  of  SNR and
  resolution for  10 different galaxies. \textbf{Bottom:} Random error
  in the age determination for the same ten galaxies }

  \label{fig:sigage}

\end{figure}

To explore the SSP fitting in more detail and to gain insight into the
effects of changing the SNR and resolution of spectra, we performed
Monte Carlo simulations for a subset of 10 galaxies.  We selected ten
random model galaxy spectra from our sample at $z=0.51$.  The template
library and the spectra were convolved to a resolution of $\Delta
\lambda=3,5,10,20$ \AA.  Noise was added to each spectrum
to give a range of signal to noise from $SNR=3,5,10,30,50,100,200$.
The SNR is here defined as the average SNR per resolution element of
the spectrum between 3000 and 9000 \AA, assuming simple shot noise.
$\chi^2$ minimization was used to find the metallicity, age and
normalization of the best fitting SSP spectrum.  For each galaxy, this
process was repeated 1000 times.  Fig. \ref{fig:sigage} shows the
average of the results from these simulations.  As an example of the
typical spread in ages, at a value of $SNR=100$ and $\Delta \lambda =
3$ \AA, the minimum and maximum values for the ten galaxies were
$\Delta_{age}=0.148-1.23$ Gyrs and $\sigma_{age}=0.025-0.045$ Gyrs .

The precision in this method, even when using the SSP, improves with
increasing SNR as expected, and this method can very reliably
reproduce the same age at high SNR for all resolutions.  The random
error on the age does decrease for smaller resolutions allowing for a
better determination of the age.  However, the poor accuracy of SSPs
is again highlighted in Fig. \ref{fig:sigage}.  The mean offset in age
seems to settle at a given value at $SNR>30$, but it does not seem to
be a monotonic function of the spectral resolution.

The accuracy of the age-dating is clearly critical for calculating
$H(z)$.  Unless systematic biases are constant with redshift and can
be subtracted out or are well behaved and can be removed with
modelling, they will contribute a significant systematic error to the
age calculation. Even in this relatively straight forward simulation
of using model galaxies from the MS, the SSPs are not able to
reproduce the age without significant bias and are thus not adequate
for the cosmic chronometer method.

\begin{table}
  \centering
  \begin{minipage}{8.5cm}
  \caption{Errors from the Age Dating}  
  \label{tab:ageerr} 
  \begin{tabular}{rrllllll}
     \hline 
     \multicolumn{1}{c}{}  & 
     \multicolumn{1}{c}{}  & 
     \multicolumn{3}{c}{SSP Fit} &
     \multicolumn{3}{c}{Model Fit} \\ 
     \multicolumn{1}{c}{z}  &
     \multicolumn{1}{c}{Age\footnote{Age of the Universe in Gyrs.}}  & 
     \multicolumn{1}{c}{Age} &
     \multicolumn{1}{c}{$\Delta$\footnote{Mean difference between measured age and mass-weighted age.}} &
     \multicolumn{1}{c}{$\sigma$\footnote{ Standard deviation of difference between measured age and mass-weighted age.}} & 
     \multicolumn{1}{c}{Age} &
     \multicolumn{1}{c}{$\Delta$$^b$} &
     \multicolumn{1}{c}{$\sigma$$^c$} \\ 
     \hline
     0.32 & 10.08& 8.15 & 0.14  & 1.85 & 8.32 & 0.01 & 0.29 \\
     0.46 & 8.95 & 7.42 & 0.20  & 1.72 & 7.22 & 0.01 & 0.33 \\
     0.51 & 8.59 & 7.06 & 0.23  & 1.65 & 6.83 & 0.02 & 0.33 \\  
     0.56 & 8.25 & 6.62 & 0.16  & 1.69 & 6.48 & 0.01 & 0.33 \\ 
     \hline
   \end{tabular}
   \end{minipage}
\end{table}

\subsection{Age-dating LRGs with model spectra}
\label{sec:spectra}

The large model library of different LRGs that we have created provide
a mapping from spectra to mass-weighted ages. We can potentially use
these spectra as our templates instead of the SSP templates.  To test
this, we extracted one spectrum from the models and then used the
remaining models to age date that spectrum.  Then, we repeated this
for all of the other spectra.  For each spectrum, we used its mass
weighted age as its fiducial age.  Testing the entire sample for 1448
$z=0.51$ model galaxies, we find a difference between the
mass-weighted age and the measured age of $\Delta_{mw} = -0.02$ Gyr
with a dispersion of $\sigma_{mw}=0.32$ Gyrs.  The result for the
other redshifts are listed in Table 2.  The systematic error,
$\Delta_{mw}$, and dispersion, $\sigma_{mw}$, have been substantially
reduced as compared to using SSPs.  In addition, these ages can be
used to measure $H(z)$ because they have very small, regular bias with
respect to the age of the Universe.

\section{An Observing Program}\label{sec:obs}

In this section, we explore the minimum observing time required to
recover $H(z)$ to a precision of $3, 5$, and $10\%$ at $z\approx
0.42$.  Table 2 indicates that uncertainties on individual ages of
galaxies could be as low as 0.3 Gyr if suitable templates can be
used. However, since it is still unclear what the uncertainty on
individual ages of galaxies will be in a realistic experiment, we
explore observing requirements for four values of this uncertainty
(0.05 Gyr, 0.5 Gyr, 1 Gyr and 2 Gyr). We consider observations at two
redshifts: $z=0.32$ and $z=0.51$, giving a redshift interval of
$\Delta z=0.19$. To simplify our estimate, the mean ages at the two
redshifts are calculated using the average age of LRGs at each
redshift (the first method discussed in \S 3). According to equation
(\ref{eq:hzerr}), the uncertainty on the mean ages will have to be
$\sigma_{<age>} = 0.03, 0.05, 0.10 $ Gyr to measure $H(z)$ with $3, 5,
10\%$ precision.

\begin{figure}
  \includegraphics[width=8.5cm,keepaspectratio]{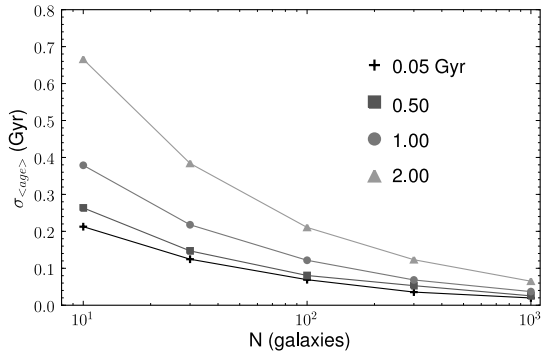} \caption{The
  uncertainty in the mean age (in Gyr) of LRGs at $z=0.51$ as a
  function of the number of galaxies used for the measurement. We plot
  curves assuming four different uncertainties (see legend) in the
  measured age of an individual galaxy in the sample.}

  \label{fig:ean}
\end{figure}

To estimate the uncertainty in mean age as a function of the number of
galaxies and the uncertainty on individual ages, we do a simple Monte
Carlo simulation. We assume the uncertainty on individual galaxy ages
are normally distributed, and the galaxy ages are drawn from the
probability distribution for galaxies at $z=0.51$ given by the fit to
equation (\ref{eq:pdf}) to the normalized age distribution.  For each
N, the simulation is repeated 1000 times and the standard deviation in
the mean age is calculated.  The results are presented in
Fig. \ref{fig:ean} and are used to calculate the total number of
galaxies required to reach our desired precision.

\begin{figure}
  \includegraphics[width=8.5cm,keepaspectratio]{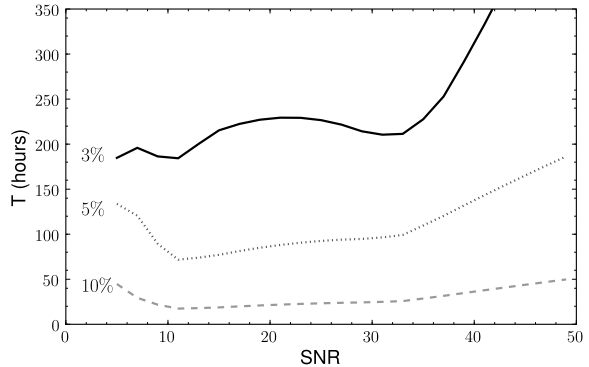}
  \caption{The total observing time required with SALT to measure
  $H(z)$ at $z=0.415$ to $3, 5,$ and $10\%$ as a function of signal to
  noise of the observations. All observations are for $\Delta \lambda
  = 3 \AA$ and an overhead of 300 s per observation.}

  \label{fig:tott}
\end{figure}

The other constraint on the total observing time is the exposure time
per galaxy.  We assume that we are only able to measure one galaxy at
a time but note that multi-object spectroscopy could be used.  In
Fig. \ref{fig:sigage}, the SNR for different spectral resolutions was
presented.  Even though the results in Fig. \ref{fig:sigage} are
obtained using SSP fitting, we use them to provide a crude estimate of
the resolution and signal to noise required to derive the random error
on the ages of individual galaxies.  In Fig. \ref{fig:tott}, we plot
the total time needed as a function of signal to noise to calculate
$H(z)$ with different precision using the Robert Stobie Spectrograph
(RSS, Kobulnicky et al. 2003) on the Southern African Large
Telescope. We estimate that a 10\% measurement is feasible using less
than 20 hours of observing time but that a 3\% measurement could
require $\sim 180$ hours, even with our fairly optimistic estimates of
$\sigma_{age}$.

We note however, that the RSS can be used in multi-object mode and,
depending on the clustering, several LRGs may be expected in each
field of view.  Using our cuts, we estimate the space density of LRGs
to be $3.9\times10^{-5}\ Mpc^{-3}$ from the SDSS (without considering
the effects of incompleteness), which is comparable to the value of
$3.5\times10^{-5} \ Mpc^{-3}$ from the MS.  If we are able to observe
two additional LRGs per set-up with $z=0.1-0.6$ while calculating
$H(z)$ to $3\%$ as outlined above, it would give us a sufficient
numbers of LRGs to calculate $H(z)$ to $10\%$ at redshifts between
$z=0.1-0.6$.  This would put a far tighter constraint on the value of
$H(z)$ than measuring it from two redshift bins alone.

\section{Conclusions}\label{sec:conc}

In this paper, we have shown that within the self-consistent Universe
of the Millennium Simulation that Luminous Red Galaxies, when selected
in an appropriate manner, can be used as cosmic chronometers.  By
selecting the galaxies from their rest-frame properties, we find that
we can create a more homogeneous sample of objects than the apparent
magnitude cuts used by Eisenstein et al. (2001).  For galaxies at
$z\sim0.3-0.6$, we find that cuts of $M_V<-23$ and $(B-V)_0>0.81$
select a sample of galaxies with similar star formation histories and
formation redshifts in the MS.

The galaxies selected in our final cut do show very similar star
formation histories with very few of the galaxies showing any star
formation since $z\sim1.7$.  The ages also show something akin to
downsizing (Cowie et al. 1996): the oldest galaxies are in the most
massive halos.  The distribution of the ages are very similar with
each showing a small tail towards younger age, but with the very
strong peak at a single age.

However we do note limitations in the use of the models of de
Lucia et al. (2006).  They are able to reproduce some of the major
trends seen in local galaxy populations and seem to have similar
number density as the SDSS out to $z\sim0.5$.  However, the model does
have some limitations in reproducing some of the observed properties
of LRGs while alternative models produce equally acceptable fits to
the data.  One such model explored here (Bower et al. 2006) was
unsatisfactory in matching the observed number density of LRGs defined
by an absolute magnitude cut at a $z\sim0.5$.

Even though these galaxies have extended SFHs, they can be used as
cosmic chronometers to recover the cosmology used in the MS.  Using
only galaxies selected from two redshifts, $H(z)$ can be calculated to
a precision of less than $3\%$ using three different methods.  All
three methods (averaging the ages, calculating a fit to the
distribution, and comparing the ages of pairs of galaxies) were able
to recover the cosmology used in de Lucia et al. (2006).

In \S \ref{sec:model}, we showed that SSPs were not sufficient to
accurately recover the ages for individual galaxies, which has also
been shown for other samples (Maraston et al. 2009, Trager \&
Somerville 2009).  Despite the relatively simple nature of the SFHs in
Fig. \ref{fig:civ}, the calculated SSP ages were dominated by the most
recent burst of star formation.  If we use the average star formation
history from the MS, we are able to replicate the properties of the
model spectra, but in the next paper, we look in far greater detail at
the modelling and fitting of LRG spectra.  However, the
semi-analytic models likely overestimate the extent of star formation
in LRGs, and SSP models may provide better fits than indicated
here. 

Finally, we estimated the required time to complete the project using
RSS on SALT.  If systematics in the age can be controlled, $H(z)$ at
$z\approx 0.42$ can be calculated to $3\%$ in a total of $\sim 180$
hours, which includes observing overheads.  It is likely that tighter
constraints could be put on the evolution of $H(z)$ with additional
data that could be obtained while making these observations. In
addition to constraining $H(z)$, it would contribute a wealth of
detailed information about the evolution of the most massive galaxies
at intermediate redshift.

Throughout this work, we have highlighted a number of assumptions that
we have made.  Even with our refined selection, the star formation
histories of these galaxies are not perfectly homogeneous as assumed
by Jimenez \& Loeb (2002), but $H(z)$ can still be calculated to an
accuracy better than $3\%$.   Other improvements on the
measurement can be made by selecting a larger or more homogeneous
sample as done by Stern et al. (2010).  Uncertainties in the star
formation history of simulated galaxies introduce far smaller errors
than the contribution associated with age-dating the galaxies. In our
next paper, we explore age-dating in more detail and note that
increased uncertainties resulting from age-dating may lead to
increased observing requirements.

\section*{Acknowledgments}

We wish to thank Bruce Bassett for the original inspiration for this
paper and for hosting the SA Cosmo Workshop, where the work begun in
earnest. We greatly appreciate the constructive criticism received from
our referee. SMC acknowledges SAAO and the NRF for support during this
project, AR and CC acknowledges the South African SKA project,
National Research Foundation (South Africa) and Centre for High
Performance Computing for financial support.  The Millennium
Simulation databases used in this paper and the web application
providing online access to them were constructed as part of the
activities of the German Astrophysical Virtual Observatory.  We
acknowledge use of the Sloan Digital Sky Survey (SDSS and SDSS-II; see
//www.sdss.org/ for funding, management, and participating
institutions).

\bsp

\label{lastpage}

\end{document}